\documentclass{appolb}
\usepackage[pdftex]{graphicx}
\pdfoutput=1

\begin{document}
\title{Isospin mixing within the symmetry restored density functional theory
and beyond\thanks{Presented at the XXXIII Mazurian Lakes Conference on Physics, Piaski, Poland, September 1-7, 2013}%
}
\author{W. Satu{\l}a$^{1}$, J. Dobaczewski$^{1,2}$, M.~Konieczka$^{1}$,
W.~Nazarewicz$^{1,3,4}$
\address{
$^{1}$Faculty of Physics, University of Warsaw, ul. Ho\.za 69, PL-00-681 Warsaw, Poland \\
$^{2}$Department of Physics, P.O. Box 35 (YFL), University of Jyv\"askyl\"a, FI-40014  Jyv\"askyl\"a, Finland \\
$^{3}$Department of Physics and   Astronomy, University of Tennessee, Knoxville, Tennessee 37996, USA \\
$^{4}$Physics Division, Oak Ridge National Laboratory, Oak Ridge, Tennessee 37831, USA}
}
\maketitle
\begin{abstract}
We present results of systematic calculations of the
isospin-symmetry-break\-ing corrections  to the superallowed
$I$=$0^+,T$=$1\rightarrow I$=$0^+,T$=$1$ $\beta$-decays, based on the
self-consistent isospin- and  angular-momentum-pro\-jec\-ted nuclear
density functional theory (DFT). We discuss theoretical uncertainties
of the formalism related to the basis truncation, parametrization of
the underlying energy density functional, and ambiguities related to
determination of Slater determinants in odd-odd nuclei. A
generalization of the  double-projected DFT model towards a
{\it no core\/} shell-model-like configuration-mixing approach is
formulated and implemented. We also discuss new opportunities in
 charge-symmetry- and charge-independence-breaking studies
offered by the newly developed DFT formalism involving
proton-neutron mixing in the particle-hole channel.
\end{abstract}
\PACS{
21.10.Hw, 
21.60.Jz, 
21.30.Fe 
}

\section{Introduction}

Isospin impurities in nuclear wave functions resulting from the
iso\-spin-symmetry non-conservation are of the order of a few
percent. Still, they impact  a plethora of  nuclear phenomena, especially in self-conjugate nuclei. Of particular importance are  the $\Delta T =0$ electric dipole transitions (E1) in
$N=Z$ nuclei, which, in the long-wavelength approximation,  are
isospin forbidden~\cite{[Tra52],[Rad52],[MDo55]}, and thus can
proceed only via the isospin mixing. This offers a method that
allows for an empirical assessment the isospin-symmetry breaking (ISB).

In case of the Fermi  and  Gamow-Teller $\beta$-decay transitions,
the selection rules are $\Delta T =0$ and $\Delta T =0, \pm 1$,
respectively, except for the $0^+ \rightarrow 0^+$ transitions that
are pure Fermi decays~\cite{[Wig39]}. Although the influence of
the ISB corrections on $\beta$-decay
rates is generally small, their precise knowledge is critically
important for the superallowed Fermi $\beta$-decays
$I$=$0^+,T$=$1\rightarrow I$=$0^+,T$=$1$ between the  isobaric analogue
states (IAS) and, to a somewhat lesser degree, for the decays between
the $T$=$1/2$ mirror nuclei. The reason is that these transitions
provide the most precise data on the vector (Fermi)
coupling constant $G_V$ and the leading  element $V_{ud}$ of the
Cabibbo-Kobayashi-Maskawa (CKM) flavour-mixing matrix. This allows
for testing the unitarity of the CKM matrix, violation of which may
indicate a {\it new physics\/} beyond the Standard Model of particle
physics, see Ref.~\cite{[Tow10a]} and references quoted therein.

Because of the smallness of isospin impurities, their accurate microscopic
calculation is a challenging task. The reason is that they result
from a subtle balance between the isospin-symmetry conserving
short-range strong interaction and the ISB long-range Coulomb interaction
that polarizes the entire nucleus. Capturing that balance is possible
 only within the {\it no core\/} approaches, which, in
heavier nuclei, reduces the possible choices to the nuclear DFT. The DFT, however, cannot be directly
used to compute  isospin impurities because of the unphysical
isospin mixing caused by the spontaneous violation of  isospin~\cite{[Eng70]}. This obstacle hindered the progress  in the field for decades. Only very recently, we have
developed the isospin- and angular-momentum projected DFT approach capable of
 treating rigorously the (conserved) rotational symmetry
and, at the same time, tackle the explicit breaking of the isospin
symmetry due to the Coulomb
field~\cite{[Sat09sa],[Sat10s],[Sat11sc],[Sat12s]}.

The aim of this work is to present a brief overview of recent results
for the isospin mixing and ISB corrections to the
$I$=$0^+,T$=$1\rightarrow I$=$0^+,T$=$1$ $\beta$-decays, focusing on
theoretical uncertainties and limitations of the employed isospin-
and angular-momentum-projected DFT framework.
All calculations presented here were obtained by using the DFT solver
{\sc hfodd} version (2.249t) or higher~\cite{[Sch12s],[Sch14]}. The paper is organized as follows.
Basic features of the
formalism are summarized in Sec.~\ref{model}. Section~\ref{deltaC} discusses the ISB corrections to
superallowed $\beta$-decays, focusing on theoretical uncertainties. A dynamic variant of the
model, which involvs the configuration mixing, is presented in
Sec.~\ref{beyond} together with preliminary applications.
Section~\ref{pnm} overviews new opportunities in studying the ISB mechanism
offered by the newly developed proton-neutron-symmetry-breaking DFT. Finally, Sec.~\ref{summy} summarizes the main findings of this work.

\section{Isospin and angular momentum projected DFT model}\label{model}

The calculation starts with solving Skyrme-Hartree-Fock (SHF)
equations without pairing using a Hamiltonian that consists the
isospin-invariant kinetic energy and Skyrme interaction, and the
 Coulomb force being the only explicit source of
isospin-symmetry violation in the model. The resulting
self-consistent Slater determinant $|\varphi \rangle$ breaks
rotational and isospin invariance. It is used as a reference
to create a basis of good-isospin-good-angular-momentum states
$|\varphi ;\, IMK;\, TT_z\rangle$:
\begin{equation}\label{ITbasis}
|\varphi ;\, IMK;\, TT_z\rangle =   \frac{1}{\sqrt{N_{\varphi;IMK;TT_z}}}
\hat P^T_{T_z,T_z} \hat P^I_{M,K} |\varphi \rangle ,
\end{equation}
where $\hat P^T_{T_z ,T_z}$ and $\hat P^I_{M,K}$ stand for the
standard isospin and angular-mo\-men\-tum projection
operators~\cite{[RS80]}, respectively. The
Hamiltonian is then rediagonalized  in the basis $|\varphi ;\, IMK;\,
TT_z\rangle$. Since the intrinsic quantum number
$K$ is not conserved and the set (\ref{ITbasis}) is overcomplete, the
rediagonalization is done by first selecting the subset of linearly
independent states, known  as {\it collective space}~\cite{[RS80]}.
For each $I$ and $T$ these states are spanned  by the so-called {\it
natural states\/} $|\varphi;\, IM;\, TT_z\rangle^{(i)}$
\cite{[Dob09ds]}.  The resulting eigenfunctions are:
\begin{equation}\label{KTmix}
|n; \,\varphi ; \,
IM; \, T_z\rangle =  \sum_{i,T\geq |T_z|}
   a^{(n;\varphi)}_{iIT} |\varphi;\, IM; TT_z\rangle^{(i)} ,
\end{equation}
where index $n$ labels the eigenstates in ascending order of
energies. The isospin impurities are defined as:
\begin{equation}
\label{truemix}
\alpha_{\rm C}^n = 1 - \sum_{i} |a^{(n;\varphi)}_{iIT}|^2,
\end{equation}
where the sum is performed for a fixed value of the isospin $T$
that dominates the wave function (\ref{KTmix}). The impurities
(\ref{truemix}) are, by construction, free from the spurious isospin
mixing.

The isospin impurities can be studied using the
isospin-only-projected variant of the approach, which is free from
singularities~\cite{[Sat09sa],[Sat10s],[Sat11sc]} plaguing
angular-momentum or particle-number
projections~\cite{[Ang01],[Rob07],[Dob07sd],[Zdu07bs],[Lac09]}. The
calculated impurities are consistent with the recent data extracted
from the giant-dipole-resonance decay studies in
$^{80}$Zr~\cite{[Cor11x]} and isospin-forbidden E1 decay in
$^{64}$Ge~\cite{[Far03]}. Both data points disagree with the standard
mean-field (MF) results, which are lower by almost 30\% due to
spurious contaminations. This agreement shows that our model is
capable of quantitatively capturing the magnitude of the isospin
mixing, which is important in the context of applying it to the
determination of the ISB corrections to the superallowed Fermi
$\beta$-decay.

\section{Isospin-symmetry-breaking corrections to superallowed $\beta$-decay}\label{deltaC}

The projected DFT method allows for a rigorous quantum-mechanical
calculation of the $I$=$0^+,T$=$1\rightarrow I$=$0^+,T$=$1$ Fermi matrix
element using the bare isospin operators $\hat T_{\pm}$, that is,
\begin{equation}\label{fermime}
|\langle I=0, T\approx 1, T_z=\pm 1;\, \varphi | \hat T_{\pm} | I=0, T\approx 1, T_z=0;\, \psi \rangle |^2
\equiv 2(1-\delta_{\rm C}).
\end{equation}
On the one hand, the state $| I$=$0, T\approx 1, T_z$=$\pm 1;\,
\varphi \rangle$ is approximated  by a double-projected state
(\ref{KTmix}), where the self-consistent Slater determinant $|\varphi
\rangle$ represents the ground state (g.s.) of even-even nucleus.
The wave function  $|\varphi \rangle$ is uniquely determined by occupying pairwise
the deformed single-particle (s.p.) orbitals  from the bottom of a
potential well up to the Fermi level. On the other hand, the double-projected state $| I$=$0, T\approx 1,
T_z$=$0;\, \psi \rangle$  represents the anti-aligned exited state of the
odd-odd $N=Z$ system. The anti-aligned configuration~\cite{[Sat12s]}
is obtained by placing the odd neutron and odd proton in the lowest
available time-reversed (or signature-reversed) s.p.\ orbitals
$|\pi\rangle\otimes |\bar{\nu }\rangle$ (or $|\bar{\pi}\rangle\otimes
|\nu \rangle$). Such arrangement manifestly breaks the isospin
symmetry. Projecting out the $T$=$1$ component of the determinant
$|\psi \rangle$ is essentially the only way of reaching the
$|T\approx 1\rangle$  configurations in $N=Z$ nuclei. Indeed, these
states are not at all representable by single
Slater determinants built by occupying unmixed
proton and neutron s.p.\ wave functions.

\begin{figure}[thb]
\centering
\includegraphics[width=0.8\textwidth,clip]{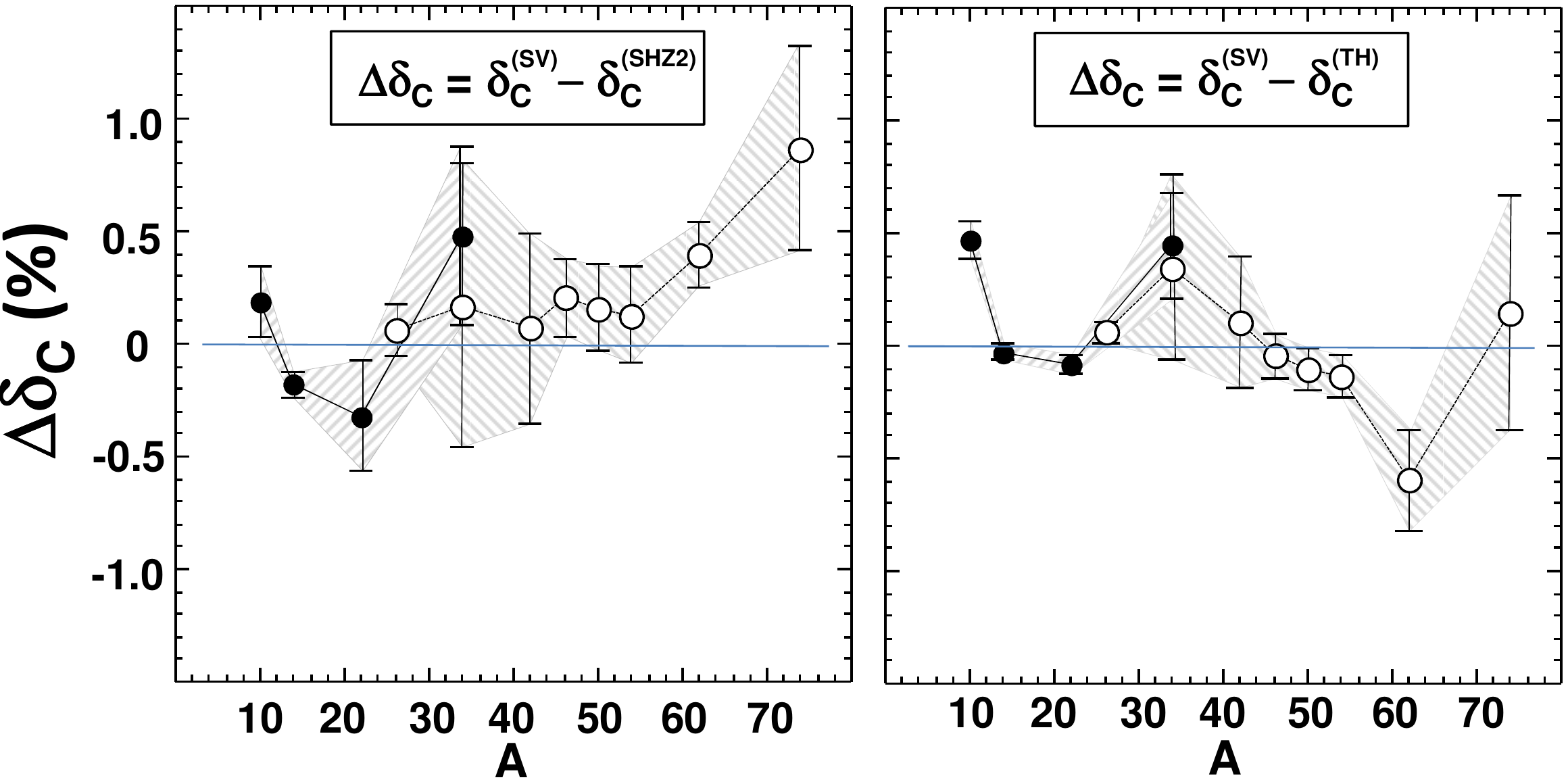}
\caption{Left: differences between the ISB corrections to
the twelve accurately measured superallowed $0^+\rightarrow 0^+$
$\beta$-transitions (excluding $A$=38) calculated using the
double-projected DFT approach with the SV and SHZ2 functionals. Right: differences between the double-projected DFT
results obtained with the SV functional and those of
Ref.~\cite{[Tow08]}. Circles and dots label the
$T_z$=$-1\rightarrow T_z$=$0$ and $T_z$=$0 \rightarrow T_z$=$1$ decays, respectively.
Shaded bands mark the estimated theoretical errors.
}
\label{Fig:delta}
\end{figure}

Calculation of matrix elements (\ref{fermime}) requires both the
isospin and angular-momentum projections~\cite{[Sat11sc],[Sat12s]}.
Since for density-dependent interactions the angular momentum
projection is ill-defined~\cite{[Ang01],[Zdu07bs]}, the method can be
used only for functionals originating from the true interaction. This
is a severe limitation that eliminates all modern density-dependent
Skyrme functionals and forces us to use the only available
density-independent parametrization SV~\cite{[Bei75s]}. Poor
spectroscopic properties of SV  result in uncertainties of
individual $\delta_{\rm C}$ corrections. To assess these
uncertainties, we developed the new density-independent Skyrme force SHZ2
by fitting its parameters to the selected bulk and s.p.\ properties
of light  doubly magic nuclei up to $^{100}$Sn, see
Ref.~\cite{[Sat12s]}. Data in light nuclei poorly constrain isovector
properties of the force. As a consequence the symmetry energy
 of SHZ2 is $a_{\rm sym} = 42.2$MeV  exceeding the
commonly acceptable value by almost 30\%. This property precludes
applications of  SHZ2  in nuclear structure studies, but it
offers an  opportunity to study the sensitivity of ISB corrections with respect to the  symmetry
energy, which governs response of the system against isovector
distortions.

Differences between the ISB corrections calculated using the
double-projected DFT with the SV and SHZ2 functionals
are shown in Fig.~\ref{Fig:delta}a. The error bars include two
sources of theoretical uncertainties: ({\it i\/}) the error due to
the basis cutoff and ({\it ii\/}) the error resulting from averaging over
the values obtained for different relative orientations of the
nuclear shapes and currents associated with the valence
neutron-proton pairs~\cite{[Sat12s]}. The error due to the basis size is rather
conservatively estimated to be of the order of $\sim$10\%. It can be, in principle,
reduced by taking a larger basis set. However, within the present
formalism, ambiguities related to the shape and current orientations
cannot be reduced further without taking into account the configuration mixing.

Surprisingly, the large difference between symmetry energies of  SV and
SHZ2  has a surprisingly modest impact on the calculated values of
$\delta_{\rm C}$. Only for the heaviest nuclei considered, the values of
$\delta_{\rm C}$ calculated using  SHZ2  are somewhat reduced as
compared to the SV results. The two sets of the calculated ISB
corrections to the $I$=$0^+,T$=$1\rightarrow I$=$0^+,T$=$1$ Fermi decay lead
to $|V_{ud}|=0.97397(27)$ and $|V_{ud}|=0.97374(27)$, for  SV and
SHZ2, respectively~\cite{[Sat12s]}. Both values result in  the
unitarity of the CKM matrix up to 0.1\% and both are fully consistent
with the result obtained by Towner and Hardy (TH)~\cite{[Tow08]}
using  different methodology based on the nuclear shell-model
combined with MF wave functions. It is
gratifying to see that also individual SV values of $\delta_{\rm C}$
are consistent within 2$\sigma$ with
the values calculated in Ref.~\cite{[Tow08]}, see
Fig.~\ref{Fig:delta}b. Two exceptions are the ISB corrections to
$^{10}$C$\rightarrow$$^{10}$B and $^{62}$Ga$\rightarrow$$^{62}$Zn
transitions. Mutually consistent DFT and TH results
are at variance with the RPA-based study of Ref.~\cite{[Lia09]}.

\section{Beyond multi-reference DFT}\label{beyond}

As discussed in the previous section, the double-projected DFT model
involves a single self-consistent Slater
determinant, representing the ground state of an even-even  $T_z$=$\pm 1$
nucleus, and a single Slater determinant $\psi$, representing the
anti-aligned configuration in an odd-odd $T_z$=$0$ system. Owing to the
ambiguities in choosing  shape and current orientations, the
latter configuration is not uniquely defined. To estimate associated uncertainties,  one can averaging  over results obtained for different
reference states~\cite{[Sat12s]}.

To overcome such difficulties, we have implemented an
extended version of the model that allows for mixing of states projected
from different self-consistent Slater determinants $\varphi_i$ representing
low-lying (multi)particle-(multi)hole excitations in a nucleus of interest.
The extension can
be viewed as a variant of the {\it no core\/} shell-model, with two-body
effective interaction (including the Coulomb force) and a
basis-truncation scheme dictated by the self-consistent deformed
Hartree-Fock solutions. The scheme proceeds as follows:
\begin{itemize}
\item
A set of low-lying (multi)particle-(multi)hole SHF states
$\{ \varphi_i \}$ is calculated along with their HF energies $e^{(HF)}_i$.
These states form a basis of reference states for a subsequent
projection.
\item
The  projection techniques are applied to the set of states $\{ \varphi_i
\}$ to calculate a family $\{ \Psi_I^{(\alpha)}, E_I^{(\alpha) }
\}$ of good-angular momentum states with  $K$-mixing
and isospin mixing treated properly. The states $\{ \Psi_I^{(\alpha)} \}$ are, in general, non-orthogonal.
\item
The mixing of states $\{ \Psi_I^{(\alpha)} \}$
is performed by solving the Hill-Wheeler equation in the
collective space spanned  by the natural states
corresponding to non-zero eigenvalues of the norm matrix, that is, by
applying the same technique which is used to handle the
$K-$mixing~\cite{[Dob09ds]}.
\end{itemize}

The model can be used to calculate spectra and transitions in
any nucleus, irrespectively of its neutron- and proton-number parities.
It can be also applied to compute $\beta$-decay matrix elements
between different nuclei. In particular,
it opens up a
possibility of detailed studies of  isovector sector of the underlying
EDFs~\cite{[Naz13s]}. The numerical stability of the
method is, however, affected by truncation errors.
Namely, the numerically unstable solutions are removed
the model space by truncating either the high-energy states $\{
\Psi_I^{(\alpha)} \}$ or the {\it natural states\/} corresponding to
small eigenvalues of the norm matrix, or by applying both truncations
simultaneously. This procedure is not fully satisfactory, but
 it is relatively reliable for energy values. Estimated errors on stable eigenvalues
rarely exceed $\pm$150keV.

The source of the obtained instabilities is not fully recognized. They
could be related to the zero range of the Skyrme force. Indeed, it is well
known that the Dirac-delta force is unstable in three dimensions and
requires regularization~\cite{[Bre47]}. Work along these lines with
finite-range density-independent EDFs~\cite{[Ben13]} is under way. Moreover,
it is not  clear whether  EDFs or two-body interactions
fitted at the MF level are realistic enough to be used in a
beyond-multi-reference DFT theory.

\begin{table}
\caption[A]{\label{tab1} The ISB corrections $\delta_{\rm
C}$ (in {\%}) in light nuclei corresponding to measured (top) and unmeasured
(bottom) superallowed $\beta$-decays in selected parent nuclei. Shown are
the results
of dynamical calculations performed in this work, $\delta_{\rm C}^{\rm (mix;SV)}$; double-projected DFT method with SV and
SHZ2 obtained in Ref.~\cite{[Sat12s]}; and the ISB corrections $\delta_{\rm C}^{\rm (TH)}$ of Ref.~\cite{[Tow08]}.
}
\begin{center}
\begin{tabular}{lrrrr}
Nucleus    &
$\delta_{\rm C}^{\rm (mix;SV)}$ & $\delta_{\rm C}^{\rm (SV)}$ &$\delta_{\rm C}^{\rm (SHZ2)}$  &
$\delta_{\rm C}^{\rm (TH)}$    \\
\hline
$T_z=-1:$ &          &          &           &            \\
$^{10}$C  & 0.668(67)&  0.65(14)&  0.462(65)&   0.175(17)\\
$^{14}$O  & 0.303(30)& 0.303(30)&  0.480(48)&   0.330(25)\\
$^{22}$Mg & 0.268(54)& 0.301(87)&  0.342(49)&   0.380(22)\\
$^{34}$Ar &  0.87(17)&  1.11(29)&   1.08(42)&   0.665(56)\\[3pt]
$T_z=0: $ &          &          &           &            \\
$^{26}$Al & 0.329(66)& 0.370(95)&  0.307(62)&   0.310(18)\\
$^{34}$Cl &  0.75(15)&  1.00(38)&   0.83(50)&   0.650(46)\\[3pt]
\hline
$T_z=-1:$ &          &          &           &            \\
$^{18}$Ne &  1.38(28)&  1.41(46)&   0.72(30)&   0.565(39)\\
$^{26}$Si & 0.427(85)&  0.47(10)&  0.529(77)&   0.435(27)\\
$^{30}$S  &  1.24(25)&  1.42(26)&   0.98(21)&   0.855(28)\\[3pt]
$T_z=0: $ &          &          &           &            \\
$^{18}$F  &  1.22(24)&  1.25(42)&   0.42(24)&            \\
$^{22}$Na & 0.257(26)&  0.35(14)&  0.216(86)&            \\
$^{30}$P  &  0.98(20)&  1.16(27)&   0.60(20)&            \\
\hline
\hline
\end{tabular}
\end{center}
\end{table}
At present, calculations can be performed only for the SV Skyrme
interaction. First results communicated in
Refs.~\cite{[Sat13s],[Naz13s]} are encouraging. Here, we present
further applications of the formalism. Table~\ref{tab1} illustrates
preliminary results for the ISB corrections in light nuclei for $A<
42$ calculated using the extended model. In these calculations, we mixed
states projected from the $|\psi^{(\rm X)}\rangle$, $|\psi^{(\rm
Y)}\rangle$, and $|\psi^{(\rm Z)}\rangle$ Slater determinants obtained
in Ref.~\cite{[Sat12s]}, which correspond to three different
orientations of the s.p.\ alignment with respect to principal axes of
the core. The results of such configuration-mixing
calculations are shown in the first column of the table, including the
error bars that are estimated to be of the order of 10\% for
numerically stable solutions and 20\% for cases requiring
regularization. The new results are fully consistent with the average
values quoted in Ref.~\cite{[Sat12s]}.

\begin{figure}[thb]
\centering
\includegraphics[angle=0,width=0.50\textwidth,clip]{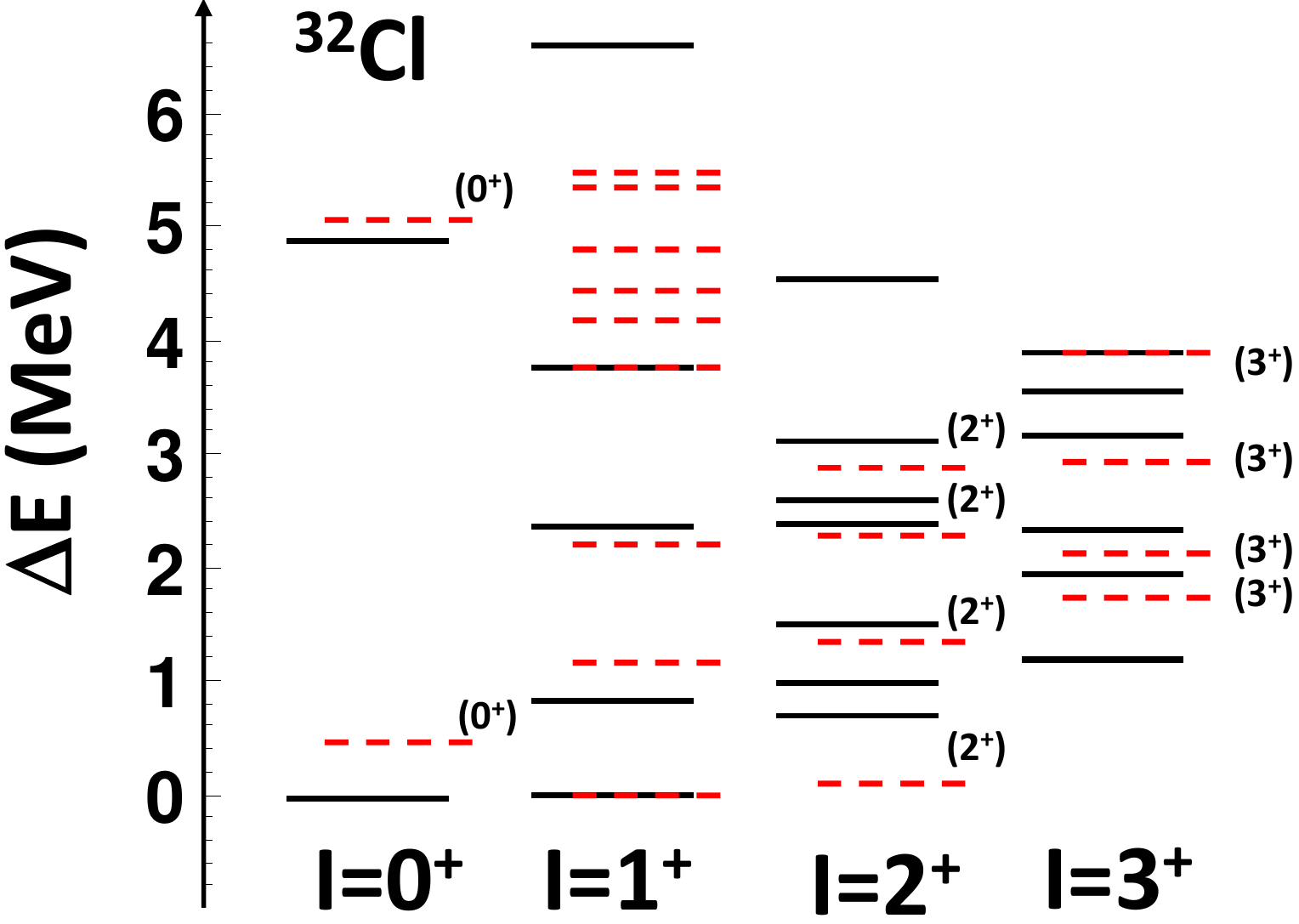}
\caption[T]{\label{Fig:32cl}
Low-spin $I$=0$^+$, 1$^+$, 2$^+$, and 3$^+$ states in odd-odd nucleus $^{32}$Cl
plotted relative to the lowest 1$^+$.
Theoretical levels are marked by solid lines. Dashed lines indicate empirical data
taken from Ref.~\cite{[Oue11]}. Note, that for all but 1$^+$ states,
the empirical spin assignments are uncertain.
}
\end{figure}
Recently, Melconian {\it et al.}~\cite{[Mel11s]} performed high
precision measurement of the $\gamma$ yields following the
$\beta$-decay of $I$=$1^+, T$=$1$ state in $^{32}$Cl to its isobaric
analogue state  in $^{32}$S reporting for the Fermi branch an anomalously large value of
$\delta_{\rm C}\approx 5.3(9)$\%. The physical
reason for this enhancement can be traced back to a near-degeneracy
of the $T$=$1$ isobaric analogue state at
7002\,keV and the $T$=$0$ state at
7190\,keV~\cite{[Oue11]}.

As discussed in Ref.~\cite{[Sat12s]}, owing to ambiguities in
choosing the reference Slater determinant, the static projected DFT
approach is not sufficient to give a reliable prediction for
$\delta_{\rm C}$. The case of $^{32}$Cl provides an excellent
 playground for testing the dynamical variant
of our model involving configuration mixing. This nucleus is a
relatively weakly bound odd-odd system with tentative spin
assignments for all but $1^+$ states, see discussion in Ref.~\cite{[Oue11]}. Figure~\ref{Fig:32cl} compares calculated
and empirical spectra of the low-spin $I$=0$^+$, 1$^+$, 2$^+$, and
3$^+$ states in $^{32}$Cl. In our calculations, nine 1p-1h
configurations were considered. We see that the level of agreement is
quite good and that the theory is capable of capturing the main features of
experiment, in particular, the placement of $I$=0$^+$ and $I$=1$^+$
states.

\begin{figure}[thb]
\centering
\includegraphics[width=0.7\textwidth,clip]{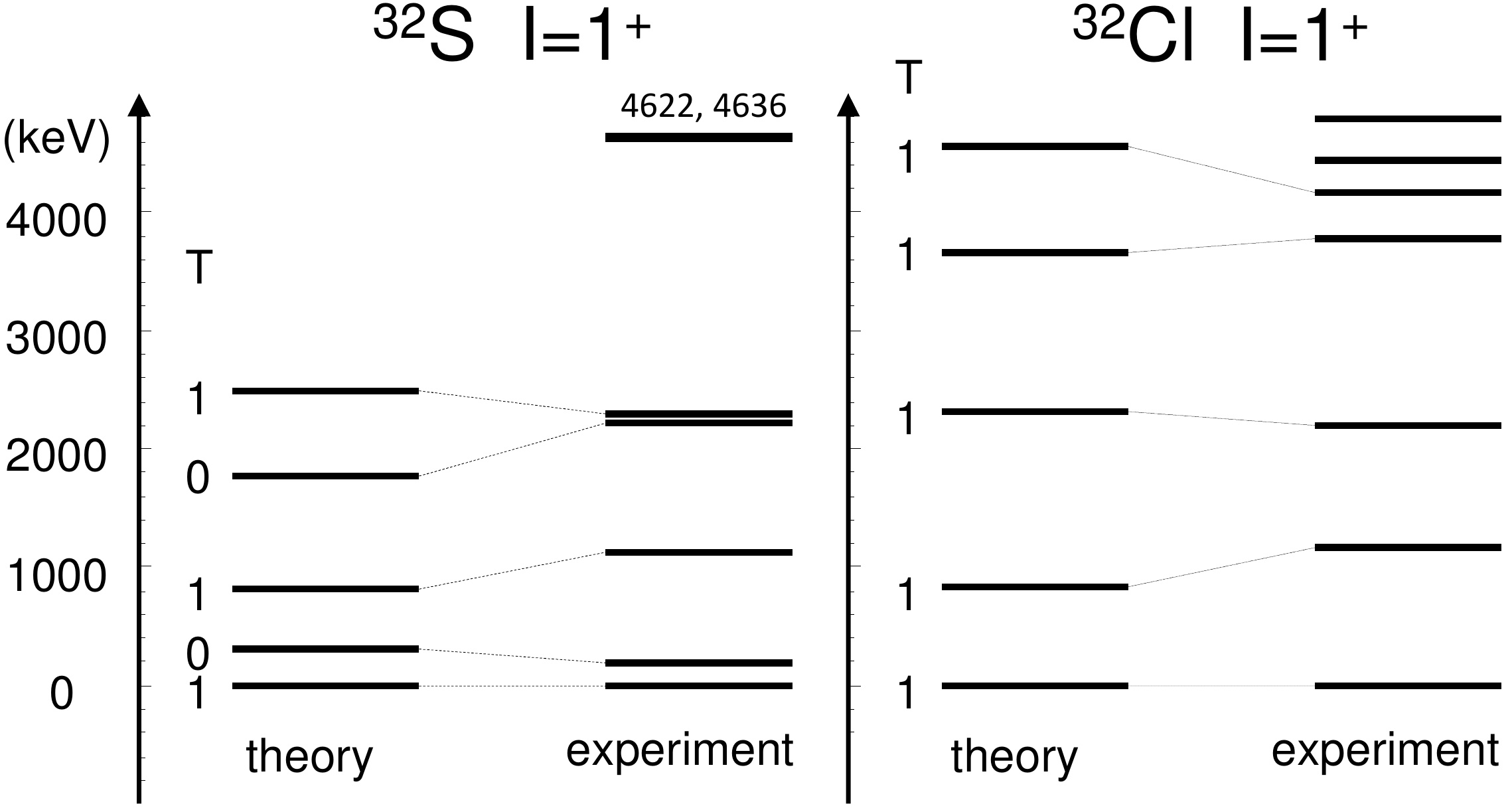}
\caption{Theoretical and empirical spin 1$^+$ states in $^{32}$S (left) and
$^{32}$Cl (right). The spectra in $^{32}$S are normalized to the isobaric
analogue 1$^+$ state and consist of states above it. In the calculations
six (nine) low-lying 1p-1h configurations were taken into account for $^{32}$S ($^{32}$Cl), respectively.
}
\label{Fig:32s-32cl}
\end{figure}

Similar dynamical calculations for $^{32}$S (with six 1p-1h configurations
included) indicate that the theory
reproduces quite well the energy splitting between the isobaric
analogue states  $I$=1$^+$,$T$=1 and $I$=1$^+$,$T$=0 --
critical for a reliable estimate of $\delta_{\rm C}$. As shown in
Fig.~\ref{Fig:32s-32cl} in our
calculations the splitting is overestimated only by $\sim$ 160\,keV and  total excitation energy of the isobaric
analogue state is underestimated by $\sim$1.3\,MeV. Furthermore, the theory well captures  the
ISB effects in the $I$=1$^+$ states in $^{32}$S and $^{32}$Cl. Indeed, the
calculated value of the ISB correction to the Fermi decay between the
isobaric analogue states is $\delta_{\rm C}\approx 6(2)$\%. Large
error bar accounts for the effect of regularization and for asymmetry
in the number and structure of the 1p-1h configurations included in
dynamical calculations in $^{32}$S and $^{32}$Cl. We stress that our calculations contain no free parameters that can be  readjusted to improve the agreement between the theory and experiment.

\section{DFT with broken neutron- and proton-number symmetries}\label{pnm}

The nucleon-nucleon (NN) strong force $V_{12}^{NN}$ is dominated by the
iso\-spin-invariant part $\sim \alpha + \beta {\vec{\tau}}^{(1)} \cdot
{\vec{\tau}}^{(2)}$. There exists, however, firm experimental evidence
that the NN  force also contains  charge independence  breaking
 (CIB) components $\sim \tau_z^{(1)} \tau_z^{(2)} $ and two types of charge
symmetry breaking (CSB) components  $\sim (
\tau_z^{(1)} + \tau_z^{(2)})$ (causing no isospin mixing) and
$\sim \alpha ( \tau_z^{(1)} - \tau_z^{(2)}) + \beta
[{\vec{\tau}}^{(1)}\times {\vec{\tau}}^{(2)}]_z $ (producing  isospin mixing) \cite{[Wil69]}. The
experimental evidence for these terms comes, among the others, from
differences in nn, pp and np phase shifts and scattering
lengths; differences in neutron/proton analyzing powers in np
scattering; binding energy differences in mirror nuclei; and  binding energy differences of isobaric analogue states.

The isospin structure of the NN force can be reexpressed in terms of
the two-body spherical tensors in isospace including the isoscalar,
isovector and isotensor components. In the DFT-rooted formalisms,
which are not directly linked to the NN interaction, one often
uses isoscalar functionals, which are bilinear (or higher order)
in isoscalar and isovector one-body densities
and currents~\cite{[Per04s],[Sat10s]}.

The dominant part of the ISB effect in the atomic nucleus is due to
the Coulomb interaction. In fact, in most of the applications
involving DFT-rooted approaches, the Coulomb force is the only term
violating the isospin symmetry. This is in spite of the perpetual problems
in reproducing the binding-energy
differences in mirror nuclei. This problem, known under
the name of Nolen-Schiffer anomaly~\cite{[Nol69]}, is rather well
studied, and it is generally agreed that its explanation requires CSB
strong interaction~\cite{[Bro00b]}, which appears to
almost exactly cancel the Coulomb exchange contribution~\cite{[Bro98],[Bro00b]}.

\begin{figure}[thb]
\centering
\includegraphics[width=0.5\textwidth,clip]{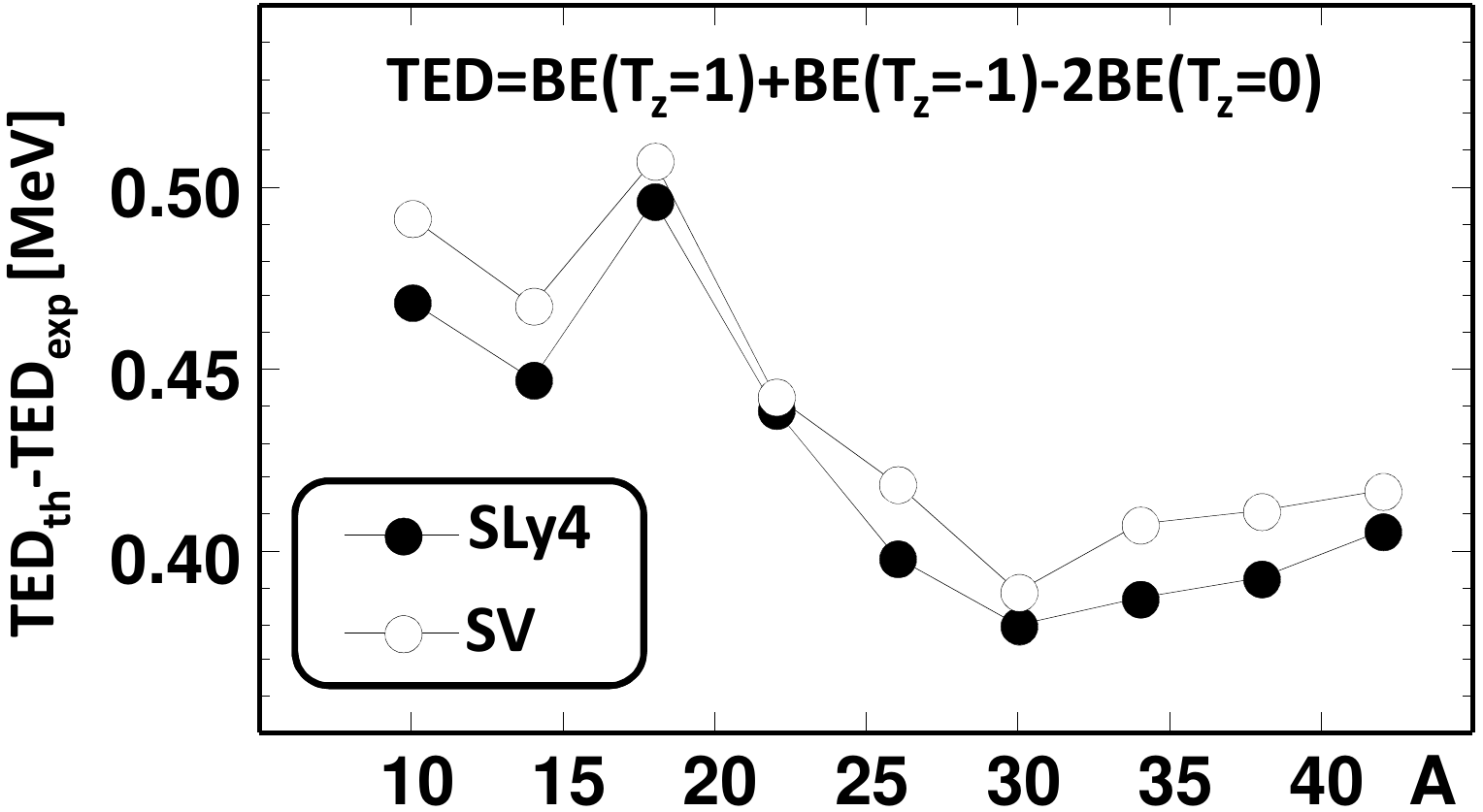}
\caption{Calculated triplet energy differences relative to
experimental data. Filled and open circles mark calculations done
with the SLy4 and SV EDFs, respectively.
}
\label{TEDs}
\end{figure}

The effect of CIB (or isotensor) interaction can be seen in
triplet binding energy differences (TED)~\cite{[Ben07a]}, which are defined as
TED$_I$$\equiv$BE$_{I,T=1,T_z=-1}$ +BE$_{I,T=1,T_z=+1}$$-$2BE$_{I,T=1,T_z=0}$.
As  mentioned in Sec.~\ref{deltaC},
the TEDs cannot be assessed by using the conventional MF,
because the $|I,T$=$1,T_z$=$0\rangle$ states are not representable by a
single Slater determinant built of unmixed proton and neutron states. Recently, we have
extended the DFT formalism by breaking the neutron-proton
symmetry at the particle-hole level~\cite{[Sat13as]}. Therein, we have
demonstrated that such a generalized DFT is capable of describing the
isobaric analogue $|T$=$1,T_z$=$0 \rangle$ states by evolving the
$|T$=$1,T_z$=$\pm 1 \rangle$ solutions by means of the tilted-axis cranking
method in isospace. This will allow us to assess the effects of  CSB and CIB
strong interactions on ISB corrections to
the superallowed $0^+ \rightarrow 0^+$ decays.

Results of our preliminary calculations for TEDs are shown in
Fig.~\ref{TEDs}. Calculations were performed by using two
different isospin-invariant Skyrme EDFs SV and
SLy4~\cite{[Cha98s]}. It is seen that the isotensor component
of the Coulomb interaction is not strong enough to explain the
experimental data. The systematic difference between the data and
calculations can be viewed as an analog of the Nolen-Schiffer anomaly
in mirror energy differences. The effect is, most likely, due to a missing CIB strong
interaction components. It is interesting to observe that the unaccounted effect $\sim$450keV
is almost $A$-  and Skyrme-force independent. Systematic studies of the TED anomaly are in progress.

\section{Summary and outlook}\label{summy}

The ISB corrections to the  $0^+\rightarrow 0^+$ superallowed
Fermi $\beta$ decays obtained  within  the isospin- and angular-momentum
projected DFT are critically overviewed. The dynamical extension of the model is
proposed that promises to cure deficiencies of the previous approach by
introducing configuration mixing. The dynamic model is applied, for the first
time, to compute: {({\it i\/})} the ISB corrections in light nuclei by mixing
states projected of the  Slater determinants in the odd-odd
$N=Z$ nuclei obtained in Ref.~\cite{[Sat12s]}; ({\it ii\/}) the
low-spin spectra in $^{32}$Cl; ({\it iii\/}) the $I$=$1^+$ states above the isobaric analogue state in $^{32}$S; and
({\it iv\/}) and the ISB correction to the $^{32}$Cl$\rightarrow$$^{32}$S Fermi $\beta$-decay branch
between the isobaric analogue $I$=$1^+$
states.  In spite of outstanding  problems related to the numerical stability of the
method, our first results are
encouraging. Indeed, our parameter-free calculations are
able to capture the main empirical features. Finally, the Nolen-Schiffer anomaly
for the isotensor CIB components of the NN interaction is formulated
and discussed. Such advanced analysis was made
possible by extending the standard nuclear DFT to the variant that
includes the mixing of proton and neutron wave functions~\cite{[Sat13s]}.

\vspace{0.3cm}
This work was supported in part by the U.S. Department of Energy under
Contract No.\ DE-FG02-96ER40963 (University of Tennessee), No.\
DE-SC0008499 (NUCLEI SciDAC Collaboration), by the Academy of Finland
and University of Jyv\"askyl\"a within the FIDIPRO programme, and by
the Polish National Science Center under Contract No.\
2012/07/B/ST2/03907. We acknowledge the CSC-IT Center for Science
Ltd, Finland for the allocation of computational resources.


\end{document}